\documentclass[aps,prl,twocolumn,preprintnumbers,
superscriptaddress,floatfix]{revtex4}

\usepackage{amssymb,multirow}
\usepackage{graphicx}

\begin{document}

\newcommand{\Tr}{\mbox{Tr\,}}
\newcommand{\beq}{\begin{equation}}
\newcommand{\eeq}{\end{equation}}
\newcommand{\bea}{\begin{eqnarray}}
\newcommand{\eea}{\end{eqnarray}}
\renewcommand{\Re}{\mbox{Re}\,}
\renewcommand{\Im}{\mbox{Im}\,}

\voffset 1cm

\newcommand\sect[1]{\emph{#1}---}



\title{Chemical Potential in the Gravity Dual of a 2+1 Dimensional System}

\author{Nick Evans and Ed Threlfall}

\affiliation{School of Physics and Astronomy, University of
Southampton,
Southampton, SO17 1BJ, UK \\
evans@phys.soton.ac.uk,ejt@phys.soton.ac.uk}

\begin{abstract}
\noindent We study probe D5 branes in D3 brane AdS$_5$ and
AdS$_5$-Schwarzschild backgrounds as a prototype dual description
of strongly coupled 2+1 dimensional quasi-particles. We introduce
a chemical potential through the U(1)$_R$ symmetry group, U(1)
baryon number and a U(1) of isospin in the multi-flavour case. We
find the appropriate D5 embeddings in each case - the embeddings
do not exhibit the spontaneous symmetry breaking that would be
needed for a superconductor. The isospin chemical potential does
induce the condensation of charged meson states.
\end{abstract}

\maketitle

\section{Introduction}
Recently there has been interest in whether the AdS/CFT
Correspondence \cite{Malda,Witten:1998qj,Gubser:1998bc} can be
used to understand  2+1 dimensional condensed matter systems (for
example
\cite{Hartnoll:2008hs,Hartnoll:2008vx,Roberts:2008ns,Hartnoll:2008kx,
Gubser:2008wv,Gubser:2008wz,O'Bannon:2008bz}). The typical UV
degrees of freedom in these systems are electrons in the presence
of a Fermi surface and a gauged U(1), QED. When brought together
in certain 2d states they can become relativistic and strongly
coupled - possibly such systems might induce superconductivity too
by breaking the gauge symmetry. The philosophy, which may be
overly naive, is to find relativistic strongly coupled systems
that show these behaviours and hope they share some universality
with the physical systems. Whether or not that linkage becomes
strong, it is interesting to study the AdS duals of 2+1d systems.

In this paper we will study the dynamics of the theory on the
world volume of a mixed D3 and D5 brane construction with a 2+1
dimensional intersection, which has previously been studied at
zero temperature in the absence of chemical potential in
\cite{Karch:2000gx,DeWolfe:2001pq,Erdmenger:2002ex}. The gravity
dual of the D3s, at zero temperature, is $AdS_5 \times S^5$, which
is dual to the 3+1 dimensional ${\cal N}=4$ super Yang Mills
theory. Here these interactions will be used to loosely represent
strongly coupled ``phonons". We will introduce 2+1d
``quasi-particles" via D5 branes (with a 2+1d intersection with
the D3s) - states connecting the two set of branes should be
expected to carry quantum numbers that interact with the D3 brane
dynamics and flavour quantum numbers associated with the number of
D5 branes - the full field theory can be found in
\cite{Erdmenger:2002ex}. We will work in the probe approximation
for the D5 branes which corresponds to quenching quasi-particle
loops in the phonon background \cite{Karch}. At zero chemical
potential the theory has ${\cal N}=4$ supersymmetry and at zero
quasi-particle mass is conformal
\cite{Karch:2000gx,DeWolfe:2001pq,Erdmenger:2002ex}. The system is
related to the higher dimensional D3-D7 intersection where the
${\cal N}=4$ gauge theory on the D3 branes has been used to
describe gluon dynamics and the D3-D7 strings quarks - some
progress in the study of the properties of mesons in 3+1d strongly
coupled gauge theories has been achieved \cite{Erdmenger:2007cm}.
The D3-D5 defect system seems a natural starting point therefore
for 2+1 dimensional systems.

To attempt to mimic a solid state system one must weakly gauge a
U(1) symmetry of the system and introduce an associated chemical
potential (by setting $A_t=\mu$). There are a number of possible
U(1)s that can play this role.

Firstly the D3-D5 world volume theory has an unbroken SO(3) global
symmetry corresponding to rotations in the 4-plane transverse to
the D5 brane.  We will introduce a chemical potential for the
quasi-particles with respect to a U(1) subgroup of the SO(3)- this
can be done by simply spinning the D5 branes in an SO(2) plane
\cite{Albash:2006bs}. The embedding of the D5 brane is described
by a scalar that is charged under this U(1) symmetry so one
naively expects to trigger superconductivity in the spirit
described in \cite{Hartnoll:2008vx} - but here we would have an
explicit understanding of the UV degrees of freedom the scalar
describes. Naively one expects the scalar describing the D5
embedding to be destabilized by the presence of a chemical
potential which gives the scalar a negative mass squared. An
equivalent statement is that one expects the centrifugal force
 associated with the rotational motion of the brane to force it off the spin axis.
In fact though we find the minimum area embedding for such
spinning probe D5 branes  and find this is not the case.

The crucial physics is that the speed of light decreases as one
moves into the centre of AdS - eventually it becomes less than the
rotation speed of the D5 brane. We show, following the higher
dimensional analysis in
\cite{Albash:2006bs,Albash:2007bq,Filev:2008xt} that there are
regular D5 embeddings into the interior which have a more
complicated embedding structure. The branes bend in the direction
of the rotation so that there are two linked scalar fields
describing the embedding - this richer theory turns out to not
include superconductivity, a subtlety on top of the arguments in
\cite{Hartnoll:2008vx}.

We can introduce mass terms for the quasi-particles that
explicitly break the U(1) symmetry and we discuss the embeddings
in these cases. There is a first order phase transition when the
R-charge chemical potential grows above the mass of the
quasi-particle bound states - below the transition the
quasi-particles exist as deconfined particles whilst above it they
are confined into bound states. This transition is analogous to
the meson melting transition seen in this system and the D3-D7
system at finite temperature
\cite{Babington,Zamaklar,Hoyos:2006z,Mateos1}. We also analyze the
finite temperature behaviour of these solutions by using the
$AdS_5$ Schwarzschild geometry as the background.

Next we study a chemical potential for the U(1) associated with
baryon number for the quark fields - this seems the most natural
candidate for how QED would manifest in the effective relativistic
theory of a solid state system. The U(1) appears in the gravity
dual as the U(1) gauge symmetry on the surface of the D5 branes -
we allow configurations with non-zero profiles for these fields on
the D5. Here we are again led by results in the D3-D7 system
\cite{Myers2006}.

At zero temperature the presence of the gauge field on the brane
naively adds in an additional constraint on the solutions that
that gauge field should be regular as the brane passes from
positive to negative values of the effective radial coordinate on
the brane. This criterion rules out small perturbations of the
standard flat embeddings of the D5 brane - instead the true
solutions become ones where the D5 brane kinks through the origin
of the space. The kink, which for large quark mass is rather
sharp, has been interpreted in \cite{Myers2006} as a tube of
strings connecting the asymptotic branes. In
\cite{Nakamura:2007nx} it has been argued that an external charge
could be responsible for the irregularity of the gauge field and that
the flat embeddings should be retained. In either interpretation,
amongst these configurations is one for zero quark mass which
turns out to simply lie straight through the origin of the space -
the chemical potential does not induce any R-charged operators to
condense. The other fields on the D5 world volume carry no net
baryon number and so do not couple to the chemical potential -
there is no condensation. The system does not therefore act like a
superconductor.

The finite temperature behaviour of these solutions with non-zero
baryon number chemical potential is also explored. The D5 brane
embeddings kink on to the black hole event horizon in this case.
As in the D3-D7 case there is a first order phase transition
between two different sorts of in-falling solutions which is
analogous to the meson melting transition at finite temperature
but zero chemical potential
\cite{Babington,Zamaklar,Hoyos:2006z,Mateos1}.

Finally we turn to isospin chemical potential in the case of
multiple (but still probe) D5 branes. This theory seems less
relevant to solid state physics because the quasi-particles have a
U(2) flavour symmetry - electrons don't! On the other hand it is
natural to discuss in this context and as advocated in
\cite{Gubser:2008wv,Roberts:2008ns} may provide some lessons for
p-wave superconductors. The embeddings of any individual brane is
simply the same as for an equal magnitude baryonic chemical
potential and there is no induced R-charge breaking at zero quark
mass. Where the theories differ is that there are vector bosons
(`$W^\pm$')on the branes' world volume that couple to the chemical
potential - we work in a truncated version of the DBI action that
is just Yang Mills theory on the D5 world volume. We show, with an
analysis very similar to that of
\cite{Gubser:2008wv,Roberts:2008ns} and recent work in the D3-D7
system \cite{Erdmenger:2008,Basu:2008}, that below some critical
value of the temperature the W-bosons condense at a second order
transition. This is dual to the formation of a spin one condensate
which is charged under U(1) isospin number in the gauge theory.
Were one to identify that U(1) with QED we would have a
superconductor.

\section{The D3 Theory}

We will represent the strong interaction dynamics with the large N
${\cal N}=4$ super Yang Mills theory on the surface of a stack of
D3 branes. It is described at zero temperature by AdS$_5\times
S^5$ \beq
\begin{array}{ccl}ds^2 & = & {(\rho^2+r^2) \over L^2} dx_{3+1}^2
\\&&\\ &&+ {L^2 \over (\rho^2+r^2)} (d\rho^2 + \rho^2 d \Omega_2^2
+ dr^2 + r^2 d \tilde{\Omega}_2^2) \label{ads4}
\end{array}\eeq where we have written the geometry to display the
directions the D3 lie in ($x_{3+1}$), those we will embed the D5
on ($x_{2+1}$, $\rho$ and $\Omega_2$) and those transverse ($r$
and $\tilde{\Omega}_2$). $L$ is the AdS radius.

At finite temperature the description is given by the
AdS-Schwarzschild black hole \beq ds^2= {u^2 \over L^2} (- h(u)
dt^2 + dx_3^2) + {L^2 \over u^2 h(u)} du^2 + L^2 d \Omega_5^2\eeq
\beq  h(u) = 1 - {u_0^4 \over u^4} \eeq

It is helpful to make the change of variables to isotropic
coordinates \beq {u \; du \over \sqrt{u^4 - u_0^4}} = {dw \over
w} \eeq and choose the integration constant such that if $u_0=0$
the zero-temperature geometry is recovered \beq  2 w^2 =
u^2 + \sqrt{u^4 - u_0^4} \eeq

The metric can now be written as \beq \label{bhm}
\begin{array}{ccc} ds^2 & = & \frac{1}{L^2} \left(
w^2+\frac{u_0^4}{4w^2}
 \right ) \left ( - \left (
\frac{w^4-\frac{u_0^4}{4}}{w^4+\frac{u_0^4}{4}} \right )^2 dt^2
+dx_3^2 \right )\\&&\\ && +\frac{L^2}{w^2} \left ( d \rho^2 +
\rho^2 d\Omega_2^2 + dr^2 +r^2 d \tilde{\Omega}_2^2 \right )
\end{array}\eeq
with $w^2 = \rho^2 + r^2$, which shares the coordinate structure
of (\ref{ads4}).

\section{Quenched Matter from a D5 Probe At T=0}

We will introduce quenched matter via a probe D5 brane. The
underlying brane configuration is as follows:
\begin{center}\begin{tabular}{ccccccccccc}
& 0 & 1 & 2 & 3 & 4 & 5 & 6 & 7 & 8 & 9  \\
D3 & - & - & - & - & $\bullet$ & $\bullet$ & $\bullet$ &
$\bullet$ & $\bullet$ & $\bullet$  \\
D5 & - & - & - & $\bullet$ & - & - & - & $\bullet$ & $\bullet$ &
$\bullet$
\end{tabular}\end{center}

In polar coordinates the D5 fills the radial direction of AdS$_5$
and is wrapped on a two sphere.

The action for the D5 is just its world volume \beq S \sim T \int
d^6\xi  \sqrt{- {\rm det} G}  \sim \int d\rho~ \rho^2 \sqrt{1 +
r^{'2}} \eeq where $T$ is the tension and we have dropped angular
factors on the two-sphere.

This is clearly minimized when $r$ is constant so the D5 lies
straight. The value of the constant is the size of the mass gap
for the quasi-particles. We will mainly be interested in the
conformal case where the constant is zero. Note the general large
$\rho$ solution is of the form \beq r = m + {c \over \rho} +..
\eeq Here $m$ is an explicit  mass term for the quasi-particles in
the Lagrangian and $c$ the expectation value for a
bi-quasi-particle operator - note $m$ has dimension one and $c$
dimension two adding to three as required for a Lagrangian term in
2+1d. The solution with non-zero $c$ is not normalizable in pure
AdS$_5$. Note that when $m=c=0$ the theory is conformal. Including
a non-zero $m$ or $c$ breaks the SO(3) symmetry  ie it breaks one
transverse SO(2) symmetry. From this it is apparent that $m$ and
$c$ carry charge under that U(1). Were $c$ to be non-zero when
$m=0$ it would be an order parameter for the spontaneous breaking
of the U(1) symmetry.

\section{R-Charge Chemical Potential/Spin}

Our theory as yet lacks the relevant perturbation of the Fermi
surface and the U(1) of QED. We will associate the U(1) with a
subgroup of the SO(3) of the $\tilde{\Omega}_2$ - for concreteness
we will use the angle in the $x_7-x_8$ directions.

To include a chemical potential we will spin the D5 brane in the
angular direction $\phi$ of this U(1) with angular speed $\mu$.

The spinning of the D5 branes implies that the quasi-particles see
a chemical potential. This is in fact a little bit of a peculiar
limit since the background D3 theory also has fields, including
scalars, charged under the U(1). We are not allowing that geometry
to backreact to the chemical potential. In fact we had better not
- the pure D3 theory has a moduli space for separating the D3s in
the transverse 6-plane. Were we to set them spinning they would
scatter to infinity since there is no central force to support
rotation. In the theory on the D3 surface there is a run away
Bose-Einstein condensation. We simply wish to switch off this
physics - it is not what we are interested in - so we forbid such
backreaction. The D3 theory is in an unstable state but will
nevertheless provide some strongly coupled interactions for the
quasi-particles that do see the chemical potential.

\subsection{An Overly Naive Ansatz}

We first look for solutions where the D5 embedding has $\phi = \mu
t$ and we will allow the position $r$ (the radial distance in
$x_7-x_8$) to be a function of $\rho$. The action is \beq S \sim
\int d\rho \; \rho^2 \sqrt{(1 + r^{'2})(1- {L^4 \over
(\rho^2+r^2)^2 } r^2 \mu^2)} \eeq Naively one is expecting the
centrifugal force from the spinning to eject the brane from the
axis at all but the end points where the boundary conditions hold
the brane. This would lead to a spontaneous symmetry breaking or
superconducting state.  We will see that this is what this naive
system tries to achieve.

The equation of motion for $r$ as a function of $\rho$ is easily
computed but unrevealing.
At large $\rho$ the solutions tend to the  no-rotation limit $r
\sim m + \frac{c}{\rho}$.

The (pair of) circle(s) in the $(\rho,r)$ plane described by $L^4
\mu^2 r^2=(\rho^2 + r^2)^2$ is clearly a zero of the action so
branes wrapped there provide a solution to the equation of motion.
Anything going within the locus described by the two circles is
moving faster than the local speed of light and is presumably not
physical. This locus is a stationary limit surface - we  call it
the ergosurface below.

There exist ``Karch-Katz" type solutions \cite{Karch} for
D5-branes that do not encounter the stationary limit surface -
these solutions essentially lie flat above everything plotted in
Fig.1. We want to know what happens to those which have a close
encounter with the ergosurface. It turns out that the curves which
minimize the action like to hit the surface at a right angle. They
then kink onto the surface where they can have zero action.

The relation between $m$ and $c$ for  curves impacting on the
stationary limit surface in this way is shown in Fig.1 - the
presence of non-zero $c$ at $m=0$ appears to indicate spontaneous
breaking of the U(1) symmetry ie superconductivity. Note there is
a first order phase transition between the Karch-Katz embeddings
and those hitting the ergosurface - we will discuss this
transition further below.

\begin{centering}
\begin{figure}[h]
\begin{centering}
\includegraphics[width=60mm]{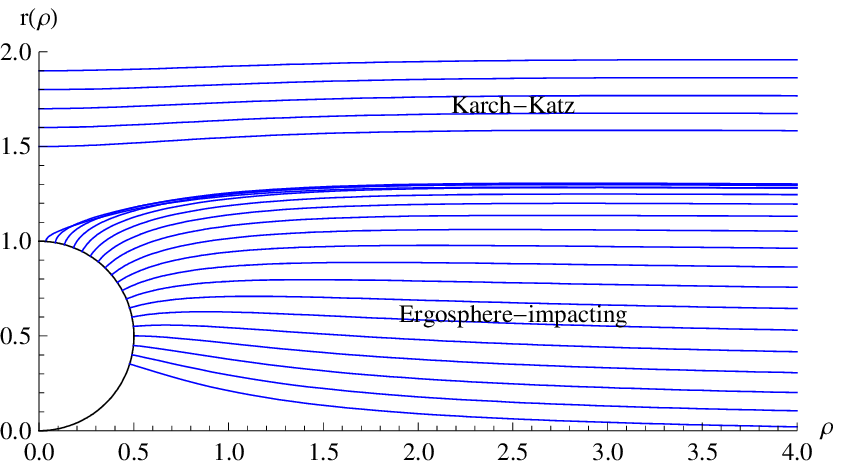} \hspace{1.0cm}
\includegraphics[width=60mm]{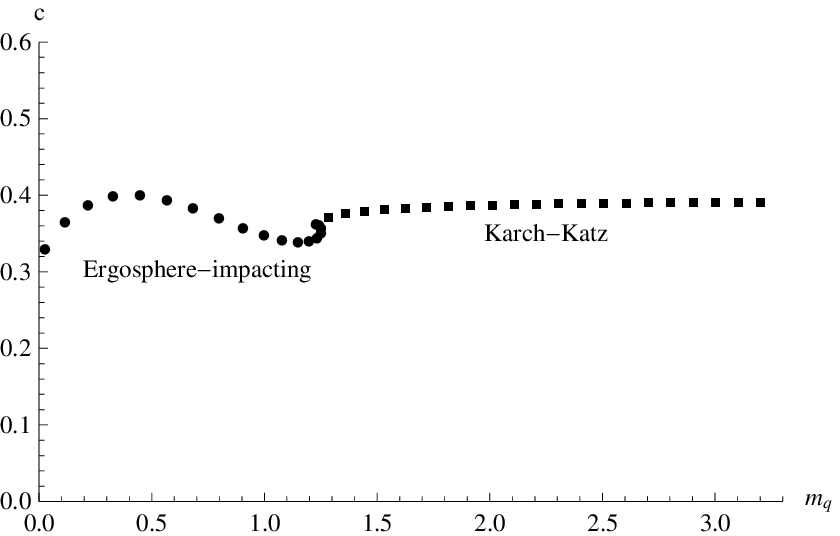}
\par
\end{centering}
\caption{Embeddings of D5 branes impacting on the ergosurface and some of the Karch-Katz type embeddings (note these actually exist down to the top of the ergosurface) (top). At the
bottom is a plot of $c$ vs $m$ for embeddings
impacting on the ergosurface (circles) and Karch-Katz embeddings (squares). The solutions oscillate around the value for the lowest
Karch-Katz solution as the D5 approaches the very top of the
ergosurface.}
\end{figure}
\par
\end{centering}

The problem of course here is that the solutions are singular at
the ergosurface where they kink. This is a sign that our ansatz is
wrong - none of this is the right physics.

\subsection{A More Sophisticated Ansatz}

We will now try a more sophisticated ansatz where the brane has in
addition some profile $\phi(w )$ where $\phi$ is the angle on
which they spin (ie $\phi = \mu t + \phi(w)$).  The ansatz is
inspired by the work in \cite{Albash:2007bq} where
similar issues are encountered when a magnetic field is switched
on on the brane's world-volume.

We find it numerically convenient to switch coordinates and write the AdS
geometry as
\begin{equation} \begin{array}{ccl}ds^2 & = & \frac{w^2}{L^2} dx_{3+1}^2+\frac{L^2}{w^2}
\left( dw^2 \right. \\ &&\\ && \left. + w^2 \left( d \theta^2 +
\sin^2 \theta d \Omega_2^2 + \cos^2 \theta d \tilde{\Omega}_2^2
\right ) \right ) \end{array}
\end{equation} the D5 will now be embedded in the $x_{2+1}, w$ and
$\Omega_2$ directions - the naive solutions above are recovered by
looking for solutions that have $\theta(w)$ and $\phi = \mu t$
where $\phi$ is the `first' angle of the $\tilde{\Omega}_2$.

In these coordinates the Lagrangian for our more ambitious ansatz
for the rotating D5 embedding is

\begin{equation} \begin{array}{ccl}
\mathcal{L} & = & w^2 \sin^2 \theta \times \\ && \\ && \sqrt{\left
( 1- \frac{L^4 \mu^2 \cos^2 \theta}{w^2} \right )\left ( 1+w^2
\theta'^2 \right )+w^2 \cos^2 \theta \; \phi'^2}
\end{array}
\end{equation}
If $\phi'\sim\mu$ the two $\mu^2$ terms compete against each other
removing the naive intuition about centrifugal force.

Since the action only depends on $\phi'$ and not $\phi$ one can
integrate the equation of motion for $\phi'$. One could then
substitute back in for $\phi'$ in terms of the integration
constant - this though gives an action with a ``zero over zero"
form at the ergosurface that is hard to work with. Instead,
following \cite{Albash:2007bq,Shock:2007}, we Legendre transform to
$\mathcal{L}' \equiv \mathcal{L}- \phi' \frac{\partial
\mathcal{L}}{\partial \phi'}$. This gives (setting $\frac{\partial
\mathcal{L}}{\partial \phi'}=J$)
\begin{equation} \begin{array}{ccc}
\mathcal{L}'&=&\frac{1}{w \cos \theta} \sqrt{\left ( 1-
\frac{L^4 \mu^2 \cos^2 \theta}{w^2} \right )}\sqrt{\left ( 1+w^2
\theta'^2 \right )}\\ &&\\ && \times  \sqrt{\left ( w^6 \sin^4
\theta \cos^2 \theta-J^2 \right )} \end{array}
\end{equation}
This has a ``zero times zero" form at the ergosurface which is
simpler to work with numerically.

For a solution that crosses the ergosurface we demand that the
action be positive everywhere and this fixes $J$ - the two terms
must pass through zero and switch signs together. Having fixed $J$
in this way one can then look at the   $\theta$ equation of motion
near the ergosurface. Expanding near the surface, and after some
algebra, one finds the following consistency equation for the
$\theta$ derivative





\begin{centering}
\begin{figure}[h]
\begin{centering}
\includegraphics[width=80mm]{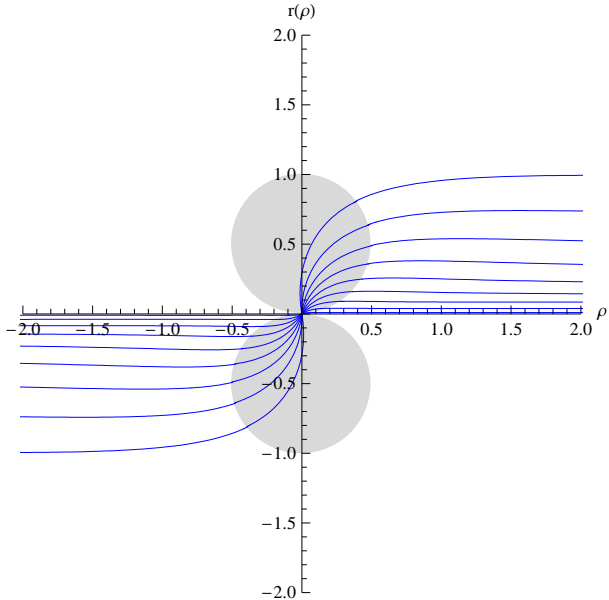}
\par
\end{centering}
\caption{A selection of solution curves for D5 embeddings.  The
grey region is the interior of the ergosurface.}
\end{figure}
\par
\end{centering}

\begin{centering}
\begin{figure}[h]
\begin{centering}
\includegraphics[width=60mm]{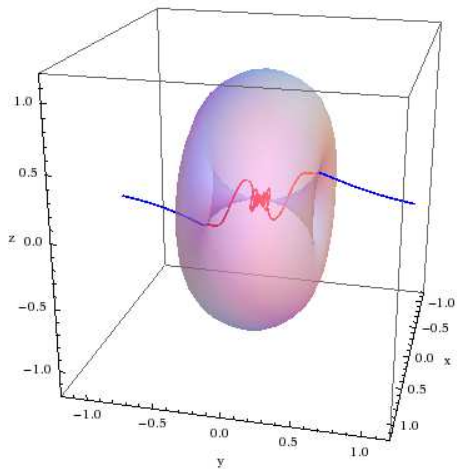}
\par
\end{centering}
\caption{A particular solution curve in the three dimensional
$(w,\theta, \phi)$ subspace. The torus represents the ergosurface.
Note the D5 rotates at speed $\mu$ in the $\phi$ direction (around
the symmetry axis).}
\end{figure}
\par
\end{centering}

\begin{equation}
w^2~ \theta'^2 + \tan \theta~ w ~\theta'-1=0
\end{equation}

There are thus two allowed gradients at the ergosurface. In
fact numerically we find choosing any gradient focuses on to the
same flow both within and outside the ergosurface. We can
numerically shoot in and out from a point near the ergosurface
in order to generate regular embeddings.

In the three-dimensional $(w, \theta, \phi)$ subspace the
ergosurface  is the torus given by $L^2 \mu \cos \theta = \pm w$,
which in a plane of constant $\phi$ gives two adjacent circles of
radius $\frac{\mu L^2}{2}$.  Fig.2 shows a sequence of regular
solutions in the $(\rho, r(\rho))$ coordinates of the previous
section. To obtain regular solutions one should make an odd
continuation to the negative quadrant as shown. We show a full D5
embedding in Fig.3 with both the $\theta(w)$ and $\phi(w)$
dependence plotted - note the D5 rotates at speed $\mu$ in the
$\phi$ direction (around the axis of the torus).

\begin{centering}
\begin{figure}[h]
\begin{centering}
\includegraphics[width=60mm]{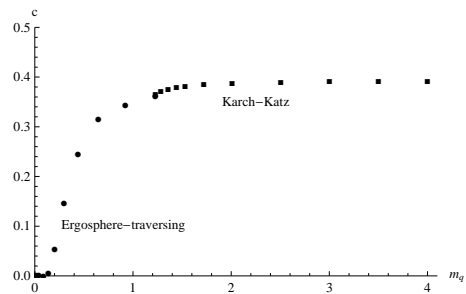}
\par
\end{centering}
\caption{A plot of $c$ vs $m$ for embeddings smoothly penetrating the ergosurface (circles) and Karch-Katz embeddings (squares). The solutions again oscillate around the value for the
lowest Karch-Katz solution as the D5 approaches the very top of
the ergosurface indicating a first order transition. }
\end{figure}
\par
\end{centering}

Clearly there is no spontaneous symmetry breaking in these
solutions - the solutions smoothly map onto the solution which
lies along the axis as the mass parameter $m$ is taken to zero. In
the field theory presumably the conformal symmetry breaking
parameter ($\mu$) which might trigger symmetry breaking is the
same parameter as that telling us there's a plasma density cutting
off the theory - there's no room for dynamics. This model turns
out not to be an example of the behaviour studied in
\cite{Hartnoll:2008vx}.

The presence of a non-trivial profile $\phi(w)$ for the embeddings
that penetrate the ergosurface indicates on the field theory side
of the duality that there is a vev for the scalar field associated
with the phase of the condensate $c$ - this would be the Goldstone
mode if there were spontaneous symmetry breaking. Note that the
regular Karch Katz embeddings, away from the ergosurface, have
$\phi(w)$ constant so there is no such vev.

Again we see there is a first order transition between the
Karch-Katz type solutions and those that enter the ergosurface
region. We plot the values of $c$ vs $m$ for these solutions in
Fig.4 - it shows the same spiral structure around the first order
transition as we saw with the naive ansatz. We will discuss the
meaning of this transition below in the thermal context.

\subsection{Thermal behaviour}

One can perform the same analysis in the thermal background.
Writing $b^4 \equiv \frac{u_0^4}{4}$, there is again a torus-like
ergosurface given by the equation
\begin{equation}
L^2 \mu \cos \theta= \pm \frac{1}{w} \frac{w^4-b^4}{\sqrt{w^4+b^4}}
\end{equation}
and also a spherical horizon at $w=b$.  One finds the horizon
always lies within the ergosurface because the local speed of
light is  zero at the horizon.  Note, below we find no phase
transition when raising the temperature through the scale of the
chemical potential. There would be a transition from a runaway
Bose-Einstein condensation to a stable theory were we to allow the
chemical potential to backreact on the geometry.

One can form the Legendre-transformed Lagrangian (which recovers the
$T=0$ case for $b=0$) \beq \begin{array}{ccc} \mathcal{L}& =&
\frac{1}{w c_{\theta}} \frac{w^4+b^4}{w^4-b^4} \sqrt{\frac{(w^4-b^4)^2}
{w^4(w^4+b^4)}\left (1-L^4 \mu^2 c_{\theta}^2 w^2
\frac{(w^4+b^4)}{(w^4-b^4)^2)} \right )} \\&&\\ && \sqrt{ \left ( 1+
w^2 \theta'^2 \right )} \sqrt{ \left ( s_{\theta}^4
c_{\theta}^2 \frac{(w^4-b^4)^2}{w^2} - J^2 \frac{w^4}{w^4+b^4}\right )}
\end{array}\eeq

\begin{centering}
\begin{figure}[h]
\begin{centering}
\includegraphics[width=60mm]{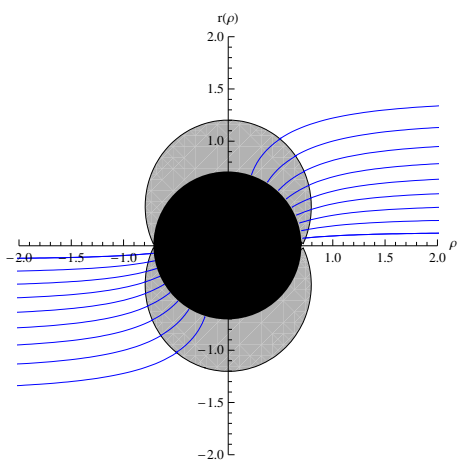}
\par
\end{centering}
\caption{A selection of solution curves for D5 branes in the
thermal geometry (with $u_0=1$).   The grey region is the interior of the
ergosurface and the black region is the interior of the event horizon.}
\end{figure}
\par
\end{centering}

The embeddings which extremize the  action fall into two types -
Karch-Katz type embeddings and those which hit the ergosurface.
Fluctuations of the former would reveal a bound state spectrum.
The latter embeddings inevitably fall onto the event horizon (a
selection of these is plotted in Fig.5 for $u_0=1$). In addition
for these embeddings that pass through the ergosurface $g_{tt}$
switches sign on the world volume - the ergosurface acts like a
horizon for the world volume fields \cite{Filev:2008xt}. Here
fluctuations would have a quasinormal spectrum along the lines of
\cite{Hoyos:2006z}.

There is therefore a first order transition in the behaviour of
the theory as the quasi-particle mass goes through the scale of
the chemical potential or temperature. This transformation is
explored in detail in \cite{Filev:2008xt}. Note here it seems the
transition is always a meson melting transition at finite
temperature.  At zero temperature the transition is driven by
quantum rather than thermal fluctuations and has been described in
terms of a metal-insulator transition in \cite{Karch:2007pd}.

\subsection{The D3-D7 System}

Much of the above parallels results already found in the D3-D7
system \cite{Albash:2006bs,Filev:2008xt}. That system describes an
${\cal N}=2$ 3+1d gauge theory with fundamental matter
hypermultiplets in the gauge background of ${\cal N}=4$ super-Yang
Mills theory. In \cite{Albash:2006bs} an analysis similar to our
``naive ansatz" was performed suggesting spontaneous symmetry
breaking. Those authors have since refined their analysis in a
related system with a background electric field
\cite{Albash:2007bq} and concluded that if regular embeddings are
insisted upon the symmetry breaking is not present (see also \cite{Shock:2007}). Were they to
update \cite{Albash:2006bs} they would find embeddings analogous
to our D5 embeddings above as they indicate in
\cite{Filev:2008xt}.

\section{Baryon number chemical potential}

Another, and perhaps the most likely, way in which to embed the
U(1) symmetry of QED into the brane set up is through the
quasi-particle number global symmetry (essentially baryon number).
The conserved vector current and its source, which is effectively
a background gauge field configuration for this symmetry, manifest
holographically as the U(1) gauge symmetry living on the world
volume of the D5 brane. We can introduce a chemical potential for
baryon number by switching on a constant $A_t$ component for this
U(1) gauge field. We will study the embeddings of such a
configuration at zero and non-zero temperature. Much of this
analysis again mirrors that for the D3/D7 system which can be
found in \cite{Myers2006}.

\subsection{Zero temperature}

The DBI action for the D5 brane including the surface gauge field
is \beq S \sim T_5 \int d^6\xi \sqrt{ \det (P[G_{ab}] + 2 \pi
\alpha' F_{ab})} \eeq

We consider embeddings of the D5 brane in the $\rho-r$ plane with
in addition
 $ 2 \pi \alpha' A_0(\rho)=A(\rho)$ to represent the chemical potential.
 The action is then of the form
\begin{equation}
\mathcal{L} \sim \rho^2 \sqrt{1+r'^2-A'^2}
\end{equation}

Since the action is independent of $A$ the equation of motion for
A implies ${\partial {\cal L} \over \partial A'}$ is a constant,
$Q$. We find \beq \label{asol} A'^2 = Q^2
\frac{1+r'^2}{\rho^4+Q^2}\eeq

It is useful to perform a Legendre transform again (${\cal L}' =
{\cal L} - A' {\partial {\cal L} \over \partial A'}$) and work
with
\begin{equation}
\mathcal{L'}= \sqrt{(1+r'^2)(\rho^4+Q^2)}
\end{equation}

The $r$ independence again gives a simple equation for the
embedding that is
\begin{equation} \label{rprime}
r'= \frac{c_1}{\sqrt{\rho^4 +Q^2-c_1^2}}
\end{equation} with $c_1$ a constant.

\begin{centering}
\begin{figure}[h]
\begin{centering}
\includegraphics[width=60mm]{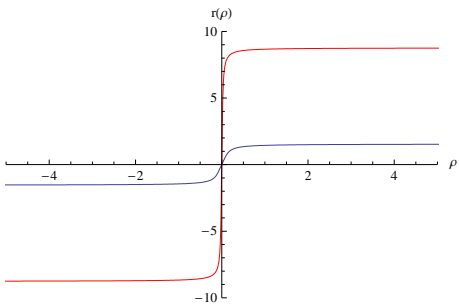}
\par
\end{centering}
\caption{Embeddings with a baryon number chemical potential for
$Q=0.1$ at $T=0$ with with nonzero $c_1$. As $c_1$ approaches the
numerical value of the charge $Q$ the quark mass can be made
arbitrarily large. }
\end{figure}
\par
\end{centering}

\begin{centering}
\begin{figure}[h]
\begin{centering}
\includegraphics[width=60mm]{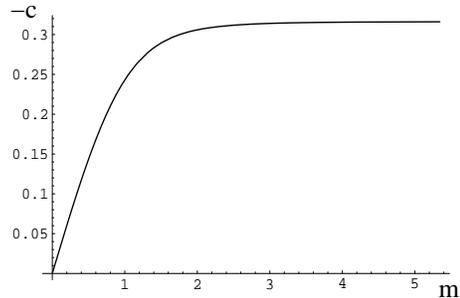}
\par
\end{centering}
\caption{Plot of parameters $c$, proportional to the condensate,
versus $m$ proportional to the quark mass in the case of baryon
chemical potential for $Q=0.1$. }
\end{figure}
\par
\end{centering}

To interpret this equation it is helpful to initially turn off the
chemical potential, $Q=0$. It is then clear that the solutions are
singular at $\rho=\sqrt{c_1}$ and the only regular case is $c_1=0$
so that $r'=0$ - we recover the usual flat embeddings.

When we allow non-zero $Q$ there become a bigger set of regular
solutions - those with $c_1 \leq Q$. These solutions are plotted
in Fig.6 for varying $c_1$ and provide an alternative
embedding, that crosses through the origin, for each value of $m$
in the large $\rho$ asymptotics of the embedding. In
\cite{Myers2006} it was argued (in the D3-D7 case) that these are
the true embeddings when there is a surface gauge field on the
brane. We can see that the flat embeddings ($c_1=0$) are not
regular from (\ref{asol})- they have a none zero gradient $A'$ at
$\rho=0$ so there will be a kink in the $A$ field as it crosses
over the $r$ axis. For the solutions that pass through the origin
though the $A$ field is regular. In \cite{Nakamura:2007nx} it has
been argued that the irregular solutions should be maintained with
the irregularity interpreted as the presence of an external
source.

As can be seen from Fig.6, whichever interpretation is taken,
the embedding for the case of massless quarks is unchanged from
the usual flat embedding. The embedding does not therefore
spontaneously break any symmetry with the introduction of a baryon
chemical potential. Away from $m=0$ for the case of the regular
embeddings the embeddings do change and there is a condensate
present. From (\ref{rprime}) we can see that up to a sign $c_1$ is
just the asymptotic parameter $c$ that determines the condensate.
For small $c_1$ the quark mass grows linearly but as $c_1$
approaches $Q$ $m$ rises sharply - the resulting plot of $c$ vs
$m$ therefore shows that the condensate asymptotes to a constant
value for large mass - see Fig.7.

Another possible source of spontaneous breaking would be if the
gauge field vev on the D5 led to other fields in the D5 brane
world volume condensing. In fact though all the fields on the D5
are in the adjoint representation of, generically, a U($N_f$)
flavour symmetry. Adjoint fields of the U(1) of baryon number are
chargeless and hence have no interaction with the gauge field.
There is no possibility for such condensation and the system is
not superconducting.

\subsection{Finite temperature}

We can also study the theory with baryon number chemical potential
at finite temperature by finding D5 embeddings with a non-zero
surface $A_t$ gauge field in the black hole geometry (\ref{bhm}).
We again set $b^4 \equiv \frac{u_0^4}{4}$.

The DBI Lagrangian for such an embedding is
\begin{equation}
\mathcal{L}= \rho^2 \frac{w^4+b^4}{w^2} \sqrt{\frac{1}{w^4}
\frac{(w^4-b^4)^2}{(w^4+b^4)} \left ( 1+r'^2 \right )-A'^2}
\end{equation}
This time we have the gauge field
\begin{equation}
A'^2 = \frac{Q^2 \frac{1}{w^4} \left (
\frac{(w^4-b^4)^2}{(w^4+b^4)}  \right ) \left ( 1+r'^2 \right
)}{\rho^4 \left ( \frac{w^4+b^4}{w^4} \right )^2+Q^2}
\end{equation}
The Legendre-transformed version of the Lagrangian is
\begin{equation}
\mathcal{L}'= \sqrt{\frac{1}{w^4} \frac{(w^4-b^4)^2}{w^4+b^4}
\left ( 1+r'^2 \right ) \left ( \rho^4 \left ( \frac{w^4+b^4}{w^4}
\right )^2+Q^2 \right )}
\end{equation}

One can shoot out from the horizon attempting to fill out the $m$
parameter space asymptotically.  The solutions are very similar to
those in the D3-D7 case as outlined in \cite{Myers2006} and see
also \cite{Karch:2007pd}. There is no spontaneous symmetry
breaking. There is a first order phase transition between the
large $m$ embeddings, that are essential flat except for a spike
down onto the black hole, and smoother embeddings that fall into
the black hole at lower $m$. This transition persists until the
chemical potential becomes large, where there is no transition and
the two phases coexist. In Fig.9 we plot the quark condensate,
$c$ versus quark mass, $m$, for varying values of Q (which
determines the chemical potential) around the critical value of Q
where the phase transition between ``spike" embeddings and smooth
horizon entering embeddings ends. The disappearance of the phase
transition is evident - the physics closely resembles that in the
D3-D7 case discussed in detail in \cite{Myers2006}.

\begin{centering}
\begin{figure}[h]
\begin{centering}
\includegraphics[width=60mm]{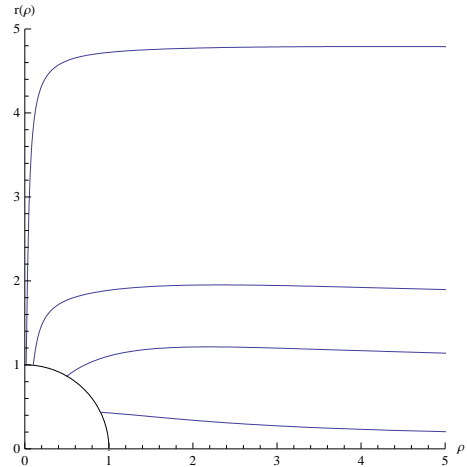}
\par
\end{centering}
\caption{Embeddings of the D5 branes for $Q=0.1$ baryon chemical
potential in units of the black hole temperature.}
\end{figure}
\par
\end{centering}

\begin{centering}
\begin{figure}[h]
\begin{centering}
\includegraphics[width=60mm]{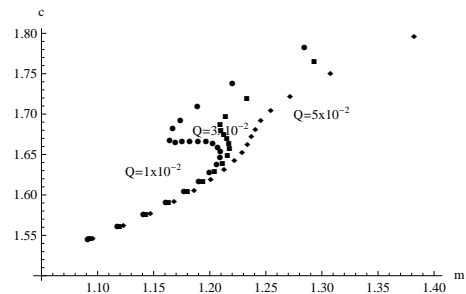}
\par
\end{centering}
\caption{Plot of the quark condensate, $c$ versus quark mass, $m$,
for varying values of Q (which determines the baryon number
chemical potential) around the critical value of Q where the phase
transition between ``spike" embeddings (the top of the s-shape)
and smooth horizon entering embeddings ends (lower part of the
s-shape).}
\end{figure}
\par
\end{centering}

\section{Isospin chemical potential}

The final possible source of a chemical potential in the D3-D5 set
up is from the isospin symmetry present when there are two or more
flavours of quasi-particle (D5) present. In contrast to the
discussion of baryon number above, there are clearly operators
which carry isospin charge eg. $\langle \bar{\psi} \gamma_0 \tau_3
\psi \rangle$. These can be expected to condense at zero isospin
chemical potential and break the symmetry spontaneously in the
spirit of the zero temperature work in \cite{Aharony:2007uu} and
the phenomenological holographic model of p-wave condensation in
\cite{Gubser:2008wv,Roberts:2008ns}. Here we will work first at
finite temperature and only consider the case of zero quark mass -
the embeddings of the branes are identical to those above for
baryon number, where one uses the modulus of the isospin as the
chemcial potential, so at zero quark mass the D5s lie straight
along the axis as shown in Figure 6. The mesons of the theory are
therefore melted by the thermal bath but the operators can
nevertheless condense. One is perhaps making a departure from any
obvious connection to a solid state system at this point since one
would require a system with a U(2) or greater flavour symmetry on
the quasi-particles - presumably there is only one sort of
electron in a solid state system!

The full theory of flavours on the D5 brane is expected to be
unstable in the presence of a chemical potential at zero
temperature. The situation is analogous to the D3-D7 system
discussed in \cite{Apreda:2005yz} - the squarks have a moduli
space in the gauge theory at zero temperature and chemcial
potential which shows up on the gravity side as a moduli space for
the size of instanton configurations on the D7 (here D5s)
world-volume \cite{Guralnik:2004ve}. An isospin chemical potential
will induce a negative mass squared for the scalars forcing the
vev or instanton size to infinity. We will simply neglect this
runaway behaviour here, fix the scalar vevs to zero and study the
fermionic operators of the theory - hopefully this tells us about
the behaviour of a theory with fermions but no scalars.

The full DBI action to all orders in the surface gauge field is
not fully known - an attempt to use the full DBI on a similar
problem in the D3-D7 set up has recently appeared
\cite{Erdmenger:2008}. We though will follow the path of the D3-D7
analysis in \cite{Basu:2008} and just use the first order
expansion of the action \beq S \sim T_5 \int d^6 \xi \sqrt{- {\rm
det} G} \left ( 1 -{1 \over 4} Tr \left ( F^2 \right ) \right ) \eeq We expect this action to
represent the dynamics well.

We will write the ansatz for the gauge fields as \beq A=
\Phi(\rho) \tau^3 dt+w(\rho) \tau^1 dx_1\eeq as in
\cite{Gubser:2008wv}. The coordinate representation of the
Yang-Mills equation is
\begin{equation}
\partial_{\rho} \left ( \sqrt{-g} F^{\rho \nu i} \right )
=-g_{YM} \left ( \delta^{im} \delta^{jl}-\delta^{il} \delta^{jm}
 \right ) A^j_{\mu} A^{\mu l} A^{\nu m}
\end{equation}

For our ansatz there are two equations of motion
\begin{eqnarray}
\left ( \sqrt{-g} g^{00} g^{rr} \Phi' \right )' &=& g_{YM} \sqrt{-g} g^{00} g^{11} w^2 \Phi\\
\left ( \sqrt{-g} g^{11} g^{rr} w' \right )' &=& g_{YM} \sqrt{-g} g^{00} g^{11} \Phi^2 w
\end{eqnarray}

Restricting ourselves to a massless quark D5 brane so the induced
brane metric is given by isotropic
AdS-Schwarzschild, the equations are (having absorbed factors of
$g_{YM}$ and $L$ into the definitions of the fields)
\begin{eqnarray}
\left ( \frac{(r^4+1)^{\frac{3}{2}}}{r^4-1} \Phi'\right )' &=& + \frac{\sqrt{r^4+1}}{r^4-1} w^2 \Phi\\
\left ( \frac{r^4-1}{\sqrt{r^4+1}} w'\right )' &=& - \frac{\sqrt{r^4+1}}{r^4-1} \Phi^2 w
\end{eqnarray}

In the isotropic coordinates one should shoot out from a small displacement $x$
from the horizon using the initial condition $w=w_0$ and $\Phi= \Phi_2 x^2$
 for constants $w_0$ and $\Phi_2$.  Near the boundary of AdS the solutions
 behave as
\begin{eqnarray} \label{isosas}
\Phi &\sim& \mu - \frac{\rho}{r}\\
w &\sim& \mu' + \frac{c}{r}
\end{eqnarray}

We search for solutions which have $\mu'=0$ because these are
solutions which are normalizable and hence describe a condensate
of the charged mesonic operator.  Requiring this to be the case
one can solve the coupled nonlinear equations to yield multiple
branches of solutions.  The lowest, monotonic branch is presumably
the stable solution and for these solutions one can plot the
dependence of the condensate on the chemical potential.  Since the
quarks we put in were massless, the only two scales are the
chemical potential and the temperature and so large chemical
potential can be equivalently viewed as low-temperature.

The results are plotted in Fig.10 - for $\mu \ll T$ there is no
condensation. For large $\mu$ though there is a second order phase
transition to a phase with a charged vector condensate. The
behaviour is similar to that previously observed in
\cite{Gubser:2008wv,Roberts:2008ns,Basu:2008,Erdmenger:2008}. As
$\mu/T$ goes to infinity the condensate $c$ tends to a finite constant ($\approx 0.3$) when measured in units of $\mu^2$, which is similar to the behaviour found in \cite{Basu:2008} for a pair of D7 probe branes - in their 3+1 dimensional theory the condensate tends to a constant in units of $\mu^3$.

\begin{centering}
\begin{figure}[h]
\begin{centering}
\includegraphics[width=80mm]{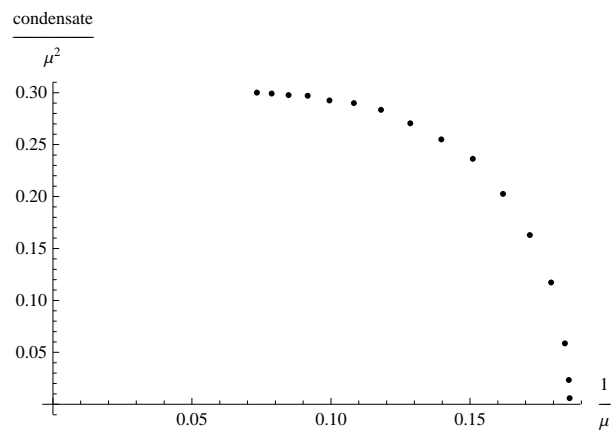}
\par
\end{centering}
\caption{Plot of the charged vector condensate (the parameter $c$
from (\ref{isosas})) in units of chemical potential squared versus $1 / \mu$ in the case of isospin
chemical potential at zero quark mass.}
\end{figure}
\par
\end{centering}

\section{Summary}

We have proposed probe D5 branes in D3 brane backgrounds as a
plausible dual for a strongly coupled quasi-particle theory in 2+1
dimensions - at zero temperature and chemical potential the theory
is supersymmetric and conformal. We introduced a chemical
potential with respect to the global U(1) symmetries associated
with R-charge and baryon number  and found the resulting regular
D5 embeddings. These embeddings do not display spontaneous
symmetry breaking and, indeed, at zero temperature and zero
intrinsic mass the theory is essentially indifferent to the
chemical potential remaining as a state of conformal
quasi-particles.

For the R-charge case we show a first order phase transition in
the massive theory as the quasi-particle mass crosses the value of
the chemical potential - on one side the quasi-particles are
confined on the other they are not. At finite temperature the
transition is between solutions that fall into the black hole and
those that don't.

At finite temperature in the baryon number case there is also a
phase transition between two different black hole embeddings which
is the equivalent of the meson melting transition at finite
temperature but zero chemical potential. If the chemical potential
becomes too large then the transition ceases to occur for any
quark mass.

Finally we introduced isospin chemical potential in the case of
two probe D5 branes and reproduced the second order transition to
a phase with a charged vector condensate previously seen in other
systems in
\cite{Gubser:2008wv,Roberts:2008ns,Basu:2008,Erdmenger:2008}

We hope that these explorations will form a useful platform from
which to find a holographic model of some real solid state system.
We note that many of the transport properties of this system have
also been recently explored in \cite{Myers:2008me}.

\vspace{1cm}

\noindent {\bf Acknowledgements:} ET would like to thank STFC for
his studentship funding. We are grateful to Rob Myers and Sean
Hartnoll for discussions at the beginning of this project. We
thank Clifford Johnson, Tameem Albash, Veselin Filev and Hsien-Hang Shieh for
guidance on their work.


\begin{thebibliography}{ll}

\bibitem{Malda}
J.~M.~Maldacena,  Adv.\ Theor.\ Math.\ Phys.\  {\bf 2}, 231 (1998)
Int.\ J.\ Theor.\ Phys.\  {\bf 38}, 1113 (1999)
[arXiv:hep-th/9711200].

\bibitem{Witten:1998qj}
  E.~Witten,
  Adv.\ Theor.\ Math.\ Phys.\  {\bf 2} (1998) 253
  [arXiv:hep-th/9802150].


\bibitem{Gubser:1998bc}
  S.~S.~Gubser, I.~R.~Klebanov and A.~M.~Polyakov,
  Phys.\ Lett.\  B {\bf 428} (1998) 105
  [arXiv:hep-th/9802109].

\bibitem{Hartnoll:2008hs}
  S.~A.~Hartnoll and C.~P.~Herzog,
  Phys.\ Rev.\  D {\bf 77} (2008) 106009
  [arXiv:0801.1693 [hep-th]].



\bibitem{Hartnoll:2008vx}
  S.~A.~Hartnoll, C.~P.~Herzog and G.~T.~Horowitz,
  arXiv:0803.3295 [hep-th].

\bibitem{Roberts:2008ns}
  M.~M.~Roberts and S.~A.~Hartnoll,
  JHEP {\bf 0808} (2008) 035
  [arXiv:0805.3898 [hep-th]].



\bibitem{Hartnoll:2008kx}
  S.~A.~Hartnoll, C.~P.~Herzog and G.~T.~Horowitz,
  JHEP {\bf 0812} (2008) 015
  [arXiv:0810.1563 [hep-th]].



\bibitem{Gubser:2008wv}
  S.~S.~Gubser and S.~S.~Pufu,
  arXiv:0805.2960 [hep-th].

\bibitem{Gubser:2008wz}
  S.~S.~Gubser and F.~D.~Rocha,
  arXiv:0807.1737 [hep-th].

\bibitem{O'Bannon:2008bz}
  A.~O'Bannon,
  arXiv:0811.0198 [hep-th].




\bibitem{Karch:2000gx}
  A.~Karch and L.~Randall,
  JHEP {\bf 0106} (2001) 063
  [arXiv:hep-th/0105132].



\bibitem{DeWolfe:2001pq}
  O.~DeWolfe, D.~Z.~Freedman and H.~Ooguri,
  Phys.\ Rev.\  D {\bf 66} (2002) 025009
  [arXiv:hep-th/0111135].

\bibitem{Erdmenger:2002ex}
  J.~Erdmenger, Z.~Guralnik and I.~Kirsch,
  Phys.\ Rev.\  D {\bf 66} (2002) 025020
  [arXiv:hep-th/0203020].



\bibitem{Karch}
  A.~Karch and E.~Katz,
  JHEP {\bf 0206}, 043 (2002)
  [arXiv:hep-th/0205236].


\bibitem{Erdmenger:2007cm}
  J.~Erdmenger, N.~Evans, I.~Kirsch and E.~Threlfall,
  Eur.\ Phys.\ J.\  A {\bf 35} (2008) 81
  [arXiv:0711.4467 [hep-th]].

\bibitem{Albash:2006bs}
  T.~Albash, V.~G.~Filev, C.~V.~Johnson and A.~Kundu,
  arXiv:hep-th/0605175.

\bibitem{Albash:2007bq}
  T.~Albash, V.~G.~Filev, C.~V.~Johnson and A.~Kundu,
  arXiv:0709.1554 [hep-th].

\bibitem{Shock:2007}
  J.~Erdmenger, R.~Meyer and J.~P.~Shock,
  JHEP {\bf 0712} 091, 2007
  [arXiv:0709.1551 [hep-th]]

\bibitem{Filev:2008xt}
  V.~G.~Filev and C.~V.~Johnson,
  arXiv:0805.1950 [hep-th]

\bibitem{Babington}
  J.~Babington, J.~Erdmenger, N.~J.~Evans, Z.~Guralnik and I.~Kirsch,
  Phys.\ Rev.\  D {\bf 69} (2004) 066007
  [arXiv:hep-th/0306018].

\bibitem{Zamaklar}
  K. Peeters, J. Sonnenschein and M. Zamaklar,
  Phys. Rev. {\bf D74} 106008, 2006
  [arXiv:hep-th/0606195]



\bibitem{Hoyos:2006z}
  C.~Hoyos, K.~Landsteiner and S.~Montero,
  JHEP {\bf 0704} 031, 2007
  [arXiv:hep-th/0612169].


\bibitem{Mateos1}
  D.~Mateos, R.~C.~Myers and R.~M.~Thomson,
  Phys.\ Rev.\ Lett.\  {\bf 97} (2006) 091601
  [arXiv:hep-th/0605046].


\bibitem{Myers2006}
  S. Kobayashi, D. Mateos, S. Matsuura, R.C. Myers, R.M. Thomson,
  JHEP {\bf 0702} (2007) 016
  [arXiv:hep-th/0611099].

\bibitem{Karch:2007pd}
  A.~Karch and A.~O'Bannon,
  JHEP {\bf 0709} (2007) 024
  [arXiv:0705.3870 [hep-th]].


\bibitem{Nakamura:2007nx}
  S.~Nakamura, Y.~Seo, S.~J.~Sin and K.~P.~Yogendran,
  Prog.\ Theor.\ Phys.\  {\bf 120} (2008) 51
  [arXiv:0708.2818 [hep-th]].



\bibitem{Erdmenger:2008}
  M. Ammon, J. Erdmenger, M. Kaminski and P. Kerner,
  [arXiv:0810.2316 [hep-th]].

\bibitem{Basu:2008}
  P. Basu, J. He, A. Mukherjee, H. Shieh,
  [arXiv:0810.3970 [hep-th]].


\bibitem{Aharony:2007uu}
  O.~Aharony, K.~Peeters, J.~Sonnenschein and M.~Zamaklar,
  JHEP {\bf 0802} (2008) 071
  [arXiv:0709.3948 [hep-th]].


\bibitem{Apreda:2005yz}
  R.~Apreda, J.~Erdmenger, N.~Evans and Z.~Guralnik,
  Phys.\ Rev.\  D {\bf 71} (2005) 126002
  [arXiv:hep-th/0504151].

\bibitem{Guralnik:2004ve}
  Z.~Guralnik, S.~Kovacs and B.~Kulik,
  JHEP {\bf 0503} (2005) 063
  [arXiv:hep-th/0405127].

\bibitem{Myers:2008me}
  R.~C.~Myers and M.~C.~Wapler,
  arXiv:0811.0480 [hep-th].




\end{thebibliography}
\end{document}